\documentstyle[12pt]{article}
\renewcommand\theequation{\thesection.\arabic{equation}}
\begin{document}
\setlength{\unitlength}{1mm}
{\hfill   hep-th/9504022,   March 1995 } \vspace*{2cm} \\
\begin{center}
{\Large\bf
 One-loop Renormalization of Black Hole Entropy Due to
Non-minimally
Coupled Matter
}
\end{center}
\begin{center}
{\large\bf  Sergey N.~Solodukhin$^{\ast}$}
\end{center}
\begin{center}
{\bf  Theoretical Physics Institute, Department of Physics,
University of
Alberta, Edmonton, Alberta, Canada T6G 2J1}
 \end{center}
\vspace*{2cm}
\begin{abstract}
The  quantum entanglement entropy of an eternal black hole is
studied.
We argue that the relevant Euclidean path integral is taken over
fields defined
on $\alpha$-fold covering of the black hole instanton. The
statement that
divergences of  the entropy are renormalized by renormalization
of
gravitational couplings in the effective action is proved for
non-minimally
coupled  scalar matter. The relationship of entanglement and
thermodynamical
entropies
is discussed.
\end{abstract}
\begin{center}
{\it PACS number(s): 04.60.+n, 12.25.+e, 97.60.Lf, 11.10.Gh}
\end{center}
\vskip 1cm
\noindent $^{ \ast}$
 Permanent address:  Bogoliubov Laboratory of Theoretical
Physics, Joint
Institute for
Nuclear Research, 141 980, Dubna, Moscow Region, Russia; \\
 e-mail: solod@cv.jinr.dubna.su
\newpage
\baselineskip=.8cm
\section{Introduction}
\setcounter{equation}0
According to the thermodynamical analogy one can apply the
laws of
thermodynamics that are valid for large statistical systems to the
description
of a single black hole \cite{Bekenstein1}. The key idea of this
approach is
that a black hole has an intrinsic entropy proportional to the
surface area of
the event horizon $\Sigma$.  This idea  has found  strong support
in Hawking's
discovery \cite{Hawking} of thermal radiation of a black hole that
allowed to
determine the entropy precisely as $S=A_\Sigma / 4G$. However,
the
microphysical explanation of the black hole entropy as  counting
of states is
still absent though many attempts have been made ( see recent
review
\cite{Bekenstein}). One of possible ways is to associate the
entropy with a
thermal bath of fields propagating just outside the horizon
\cite{a1}.
Recently, it also has been proposed to treat black hole entropy as
an
entanglement entropy \cite{Srednicki}, \cite{Sorkin}. Starting with
the pure
vacuum state one traces over modes of quantum field
propagating inside the
horizon and obtains the density matrix $\rho$.  The entropy then
is  defined
by the standard formula
$S=-Tr \rho \ln \rho$. It measures  the number of inside modes
which are
considered
as internal  degrees of freedom of the hole. In a similar manner,
Frolov and
Novikov
\cite{Frol.} suggested  to trace over modes outside the horizon.

The  calculations for the Rindler space and black holes
\cite{CW}-\cite{Cognola}  have shown  that entropy is divergent .
This is
essentially due to the short-distance correlations between the
inside and
outside modes.

The purpose of this paper is to demonstrate, following the
previous
investigations \cite{Sus}, \cite{Jac}, \cite{Sol},  that this
divergence is
really the ultraviolet one typically  appearing in quantum field
theory and it
can be removed by standard renormalization of the gravitational
couplings in
the effective action.
In  the earlier paper \cite{FS} we  have given  a proof of this for  a
minimally
coupled scalar field and noted that the non-minimal  coupling
needs a
special  consideration. The reason  of this is that there exists a
$\delta$-like
potential in the  field operator  due to the scalar curvature $R$
behaving as
a distribution on manifold with conical singularity. Below we
consider this in
more details. We start  in Section 2. with formulating the
Euclidean path
integral which is relevant  to calculation of the entanglement
entropy
of a black hole. In Section 3. we   formulate the statement about
renormalization of  black hole entropy.  The proof of it
 for non-minimally
coupled scalar field is given in Section 4. We conclude in Section
5. with some
remarks concerning the relationship of the entanglement
(statistical)
and thermodynamical entropies. The Appendixes  A  and B
contain      basic
formulas for curvature tensors  and heat kernel expansion on
manifolds
with conical singularities obtained in previous study.

\section{Euclidean path integral  for entanglement entropy.}
\setcounter{equation}0
The horizon surface $\Sigma$ naturally separates the whole
space-time of a
black hole on  the regions  $R_+$ and $R_-$ the free information
exchange
between which is impossible. This is obviously due to the fact
that the global
Killing vector $\xi_t=\partial_t$, generating translations in time,
becomes
null,
$\xi^2_t=0$, on the horizon. Therefore, any light signal emitted
from any point
 on  the horizon or behind it, never can reach an outside observer.
So events
happening in the part of the space-time beyond the horizon are
unobservable
for him in principle.
This concerns excitations  of quantum fields as well. They  are
naturally
separated on 'visible'
(propagating in the region $R_+$) and 'invisible'' (propagating in
the region
$R_-$) modes.
The partial loss of information about  the microstates  composing
the concrete
macrostate  typically appears in statistical description of
systems with large
number of degrees of freedom. We can see that  a similar
phenomenon naturally
happens for a black hole. This fact certainly lies in the principles
of the
thermodynamical analogue allowing to apply laws of
thermodynamics to a hole.

\bigskip

The situation , when a part of states of the system  is unknown, in
quantum
mechanics is usually described by  density matrix. Assume, that
the quantum
field $\phi$, being considered  on the whole space-time, is in
pure ground
state

\begin{equation}
\Psi_0=\Psi_0 (\phi_+,\phi_-)
\label{2.1}
\end{equation}
which is  the function of both visible ($\phi_+$) and invisible
($\phi_-$)
modes.
For an outside observer it is in the mixed state described by  the
density
matrix

\begin{equation}
\rho (\phi_+^1, \phi_+^2)=\int{} [ {\cal D} \phi_-] \Psi^+_0
(\phi_+^1, \phi_-) \Psi_0 (\phi^2_+, \phi_-)
\label{2.2}
\end{equation}
where one traces over all invisible modes $\phi_-$.
Then the entropy defined as

\begin{equation}
S_{geom}=-Tr  \hat{\rho}  \ln \hat{\rho}, \   \ \hat{\rho}={\rho \over
Tr
\rho}
\label{2.3}
\end{equation}
is so-called entanglement (or geometric) entropy
\cite{Srednicki}-\cite{Frol.}.

\bigskip

Applying this construction to a black hole, we identify  all  the
invisible
modes
with internal  degrees of freedom and  (\ref{2.3})
with entropy of the  hole.  The ground state of the black hole is
given by  the
Euclidean functional integral  \cite{BFZ} over fields defined on
manifold $E'$
which
is half-period part of the black hole instanton with metric

\begin{equation}
ds^2_{E'}=\beta^2_H g(\rho) d\varphi^2 +d\rho^2 +r^2(\rho)
d^2\Omega
\label{2.4}
\end{equation}
where angle variable $\varphi$  lies in the interval  $-{\pi \over 2}
\leq
\varphi  \leq {\pi \over  2}$.  The inverse Hawking temperature
$\beta_H$ is
determined by  the derivative of  the metric  function $g(\rho)$ on
the horizon
($g(\rho_h)=0$),
$\beta_H={2 \over g'(\rho_h)}$.
The $\phi_+$ and $\phi_-$ which enter as arguments in (\ref{2.1})
are the fixed
values at the boundaries $\phi_+=\phi (\varphi={\pi \over 2})$;
$\phi_-=\phi
(\varphi=-{\pi \over 2})$, giving the boundary condition in the path
integral.
The density matrix $\rho (\phi^1_+,
\phi^2_+)$ obtained by tracing $\phi_-$-modes is defined by the
path integral
over fields on the full black hole instanton $E \  (-{3 \over
2}\pi\leq\varphi\leq {1 \over 2}\pi)$ with cut along the $\varphi={\pi
\over
2}$
axis and taking values $\phi^{1,2}_+$ above and below the cut.
The trace $Tr \rho$ is obtained by equating the fields across the
cut and
doing the unrestricted Euclidean path integral on the complete
black hole
instanton $E$. Analogously, $Tr \rho^n$ is given by  the path
integral over
fields defined on $E_n$, $n$-fold cover of $E$.  Thus,  $E_n$ is
the manifold
with abelian isometry (with respect to angle rotation  ${\partial
}_\varphi$)
with
horizon  surface $\Sigma$ as stationary set.  Near $\Sigma$ the
$E_n$ looks as
direct product $E_n=\Sigma \otimes {\cal C}_n$, where ${\cal
C}_n$ is
two-dimensional cone with angle deficit $\delta=2\pi(1-n)$. This
construction
can be analytically continued to
arbitrary (not integer) $n \rightarrow  \alpha={\beta \over
\beta_H}$.

Define now the partition function

\begin{equation}
Z(\beta)= Tr \rho^\alpha
\label{2.5}
\end{equation}
which is path integral over fields defined on $E_\alpha$, the
$\alpha$-sheeted
covering of $E$. Then the geometric entropy (\ref{2.3})
takes the standard thermodynamical form

\begin{equation}
S_{geom}=- Tr (\hat{\rho} \ln \hat{\rho})= (-\alpha \partial_\alpha
+1)
\ln Z(\alpha) |_{\alpha=1}
\label{2.6}
\end{equation}
being expressed via partition function $Z$.  We see that $\beta$
plays the role
of the inverse temperature. After all calculations one must put
$\beta=\beta_H$
in (\ref{2.6}).
Assuming that the dynamics of matter fields is determined by a
differential
operator
$\hat{\Delta}$ we obtain that the relevant partition function
(\ref{2.5}) is
given by  the determinant

\begin{equation}
Z(\beta)=\det{}^{-1/2} \hat{\Delta}
\label{2.7}
\end{equation}
considered on  $E_\alpha$. It is essential that $E_\alpha$ is
manifold with
conical singularity since namely the singularity produces in the
effective
action $W(\alpha)=- \ln Z(\alpha)$ terms
proportional to ($1-\alpha$)  that contribute to the entropy
(\ref{2.6}).

One can see that the partition function (\ref{2.5}) looks as a
thermal one

\begin{equation}
Z(\beta)= Tr e^{-\beta \hat{H}}
\label{2.8}
\end{equation}
with $\beta$ playing the role of inverse temperature, $\hat{H}$
being a
relevant Hamiltonian. This fact was previously observed in
\cite{KS}
for the Rindler space and was  supposed to be  general.
The relevant Euclidean path integral  for the entanglement
entropy of the
Rindler space was found in \cite{CW}. The exact construction of
the wave
function of a black hole was  proposed in \cite{BFZ}.  The
thermality  of the
corresponding density matrix was established in \cite{Jac1}.

\bigskip

\section{The statement.}
\setcounter{equation}0
As defined in  previous section  entanglement entropy is not free
from
the ultraviolet divergences. They result  from the corresponding
divergences of
the  effective action
$W(\alpha)$. It was shown for the minimally coupled scalar matter
\cite{a4}  that
the divergent part of  the effective action on $E_\alpha$ is really
sum of
volume and
surface terms:

\begin{equation}
W_{div} (\alpha)=W_{div}^{vol} +W_{div}^{surf}
\label{3.1}
\end{equation}
The volume  term in (\ref{3.1}) is standard one.  It is proportional
to
$\alpha$ not contributing to entropy. The second
term is given by integral  over   the singular surface $\Sigma$. It
is
proportional to $(1-\alpha)$ and hence contributes to the entropy
resulting to
its divergence \cite{Sol1}. The origin of these divergences lies
obviously in
the short-distance correlation between 'visible' and 'invisible'
modes which is
concentrated  at the surface   $\Sigma$ separating regions
$R_+$ and $R_-$.

However, it was  proposed  in number of papers that the
divergences of entropy
can be removed by the standard renormalization  of the
gravitational couplings. Indeed, the higher curvature terms are
necessarily
generated by quantum corrections . Therefore, such a $R^2$
terms must be added
from the very beginning with some bare constants
($c_{1,B}, c_ {2,B}, c_{3,B}$) (tree-level) to absorb  the one-loop
infinities. The bare (tree-level) gravitational functional thus takes
the
form\footnote{Of course, we assume an addition to (\ref{3.2}) due
to a
classical matter which can be in principle rather complicated.}

\begin{equation}
W_{gr}=\int{}\sqrt{g}d^4x \left( -{1 \over 16\pi G_B} R
+c_{1,B}R^2+c_{2,B}
R^2_{\mu \nu} +c_{3,B} R^2_{\mu\nu\alpha\beta} \right)
\label{3.2}
\end{equation}
The corresponding tree-level entropy can be obtained within the
procedure
considered in the previous section as replica of the action
(\ref{3.2}) on
introducing of the conical singularity. The conical singularity at
the horizon
$\Sigma$ manifests itself in that a part of an curvature tensor for
such a
manifold $E_\alpha$ behaves as a distribution having support on
the surface
$\Sigma$  \cite{Teitelboim}, \cite{FS2} (see Appendix A). Hence,
the action
(\ref{3.2}) being considered on  $E_\alpha$ has volume  and
surface terms:

\begin{equation}
W_{gr}[ E_\alpha ]=W_{gr}^{vol}[E_\alpha /  \Sigma ]+
W^{surf}_{gr} [\Sigma]
\label{3.3}
\end{equation}
where the volume part is given by integral (\ref{3.2}) over  regular
part of
the manifold $E_\alpha$. This part is obviously proportional to
$W_{gr}^{vol} \propto \alpha$. So the whole contribution  to the
entropy comes
from the surface term. Using formulas of Appendix A
((\ref{A1})-(\ref{A4})) we obtain finally for the tree-level entropy
\cite{FS},
\cite{FS2}:

\begin{equation}
S(G_B, c_{i,B})= {1 \over 4G_B} A_\Sigma-\int_{\Sigma} \left( 8\pi
c_{1,B} R
+4\pi c_{2,B}R_{ii}+8\pi  c_{3,B}R_{ijij} \right)
\label{3.4}
\end{equation}
We see that the classical law $S={1 \over  4G}A_\Sigma$ gets
modified due to
$R^2$-terms in the action  (\ref{3.2}). The additional term  now
depends both
on  external and internal geometries of the surface $\Sigma$.
It should be noted that (\ref{3.4}) exactly coincides with entropy
computed by
the Noether charge method of Wald \cite{Wald}.

\bigskip

The main point now is that the divergent part of the entanglement
entropy
(\ref{2.6})  is such  that  its sum with the tree-level entropy
(\ref{3.4})

\begin{equation}
S(G_B, c_{i,B})+S_{div}(\epsilon)=S(G_{ren}, c_{i, ren})
\label{3.5}
\end{equation}
takes again the tree-level form $S(G_{ren}, c_{i,  ren})$ expressed
through
the renormalized constants $G_{ren},     \  c_{i,  ren}$. They are
related with
the bare
constants by usual equations originated from the one-loop
renormalization of
the action

\begin{equation}
W_{gr}(G_B, c_{i,  B})+W_{div}(\epsilon)=W_{gr} (G_{ren}, c_{i,
ren})
\label{3.6}
\end{equation}
being considered on regular  space-times without horizons.

Thus, divergences of the entanglement entropy are removed by
the standard renormalization of the  gravitational couplings. So,
no special
renormalization procedure for  entropy is required.

This statement for  the Newton constant $G$ has  been advocated
in \cite{Sus},
\cite{Jac} when considered divergences of the entropy of the
Rindler
space-time.
Necessity to renormalize also the higher curvature couplings was
argued in
\cite{Sol} for the entropy of the Schwarzshild black hole. For
minimal coupling
this statement was proved   in  \cite{FS} for general black hole
metric.
In recent preprint \cite{Myers} this procedure was  checked for
the
 Reissner-Nordstr\"{o}m black hole. Below we demonstrate this
statement
 for the non-minimally coupled scalar  matter generalizing the
result of
\cite{FS}.

\bigskip

\section{The  heat kernel expansion.}
\setcounter{equation}0
For non-minimally coupled scalar field the curvature directly
enters  into the
matter action:

\begin{equation}
W_{mat}={1 \over 2}\int{}[(\nabla \phi)^2 + \xi R \phi^2]
\label{4.1}
\end{equation}
Considering (\ref{4.1}) on manifold $E_\alpha$ we must take into
account the
$\delta$-like contribution of  the curvature coming from the
conical
singularity (see (\ref{A0})) \cite{Teitelboim}, \cite{FS2}:

\begin{equation}
R=\bar{R} + 4\pi(1-\alpha) \delta_\Sigma
\label{4.2}
\end{equation}
where $\bar{R}$ is  the regular part of the scalar curvature.
Therefore, the
quantization of non-minimal matter on $E_\alpha$ forces with the
problem of
treating operators with $\delta$-like potential.
Applying (\ref{4.2}) to the action (\ref{4.1}) we obtain that

\begin{equation}
W_{mat}=2\pi(1-\alpha)\xi\int_{\Sigma} \phi^2+{1 \over 2}
\int_{E_\alpha} \phi
(-\Box_\xi)\phi
\label{4.3}
\end{equation}
where we denote $\Box_\xi=\Box -\xi \bar{R}$ and assume
regularity of the field
$\phi$ on the singular surface $\Sigma$.

Then, considering the path integral over  the scalar field $\phi$ we
get

\begin{equation}
Z=\int{}[{\cal D} \phi ] e^{-2\pi  (1-\alpha)\int_{\Sigma} \phi^2}
e^{-{1 \over 2} \int_{E_\alpha} \phi (-\Box_\xi) \phi}
\label{4.4}
\end{equation}
Expanding\footnote{We proceed the perturbation expansion with
respect to
$(1-\alpha )$. The first term of the expansion is well-defined  (see
(\ref{4.11})). The next terms, however, are expected to be
ill-defined due to
contributions like $\delta^2 (0)$. The indication of this can be
found in
\cite{Allen}. In principle, we could use some type of regularization
similar to
that of  \cite{FS2} to give a sense to such a terms. It should be
noted,
however, that these terms are irrelevant  for the calculation of
entropy.
I thank D.Fursaev for this remark.} the first factor in (\ref{4.4})
by powers of $(1-\alpha)$ and omitting higher terms we have

\begin{equation}
Z=\bar{Z} \left( 1-2\pi \xi (1-\alpha) < \int_{\Sigma} \phi^2 >_{\bar{Z}
}\right)
\label{4.5}
\end{equation}
where  the average $<  \  >_{\bar{Z}}$ is taken with respect to
measure defined
by functional integral

\begin{equation}
\bar{Z}=\int{}[{\cal D} \phi ]e^{-{1 \over 2} \int_{E_\alpha} \phi
(-\Box_\xi)
\phi}
\label{4.6}
\end{equation}
Equivalently,  this can be written as follows

\begin{equation}
\ln{Z}=\ln{\bar{Z}} -2\pi\xi(1-\alpha) <\int_{\Sigma} \phi^2
>_{\bar{Z}}
\label{4.7}
\end{equation}

For $\ln{\bar{Z}}$  the following heat kernel expansion is known
\cite{a4}:

\begin{eqnarray}
&&\ln{\bar{Z}}=-{1 \over 2} \ln det (-\Box_\xi )={1 \over 2}
\int_{\epsilon^2}^{\infty} {ds \over s}  Tr \bar {K}_{E_\alpha} (s),
\nonumber \\
&&\bar{K}_{E_\alpha} (s)=e^{-s \Box_\xi}={1 \over (4\pi s)^{d \over
2}}
\sum_{n}^{}{} \bar{a}_n s^n, \ \ s \rightarrow 0
\label{4.8}
\end{eqnarray}
where the coefficients $\bar{a}_n (x,x)$ generally take the form

\begin{equation}
\bar{a}_n(x,x)=\bar{a}^{st}_n (x,x)+ \bar{a}_{n, \alpha} (x,x)
\delta_\Sigma
\label{4.9}
\end{equation}
The $\bar{a}_n^{st}(x,x)$ are standard  \cite{BD}  heat kernel
coefficients
given by  the local functions of curvature tensors (see Appendix
B). The second
term in (\ref{4.9})
has support only on the singular surface $\Sigma$,
$\bar{a}_{n,\alpha}(x,x)$
is  a local function of projections of a curvature tensors on  the
subspace
normal
to $\Sigma$. The exact form of coefficients $\bar{a}_{n,\alpha}
(x,x)$
is given in Appendix B.

On the other hand, by standard arguments we have

\begin{equation}
< \phi (x) \phi (x') >= \Box^{-1}_\xi=\int_{\epsilon^2}^{\infty} ds
e^{-s
\Box_\xi}
\label{4.10}
\end{equation}
Inserting  (\ref{4.8})-(\ref{4.10}) into (\ref{4.7}) we finally obtain

\begin{eqnarray}
&&\ln Z= {1 \over 2} \int_{\epsilon^2}^{\infty} {ds \over  s}  Tr
K_{E_\alpha}
(s), \nonumber \\
&&Tr K_{E_\alpha} (s)= Tr \bar{K}_{E_\alpha} (s) -4\pi\xi
(1-\alpha)s Tr_\Sigma
\bar{K}_{E_\alpha} (s)
\label{4.11}
\end{eqnarray}
where  the $x$-integration in $Tr_\Sigma$  is taken only over the
surface
$\Sigma$.
Identity (\ref{4.11}) allows us to write  the following expansion for
the heat
kernel $K_{E_\alpha}(s)$:

\begin{eqnarray}
&&Tr K_{E_\alpha} (s)= {1 \over (4\pi s)^{d/2} } \sum_{n}a_n s^n,
\nonumber \\
&&a_n=\int_{E_\alpha} \bar{a}_n (x,s)- 4\pi \xi (1-\alpha)
\int_{\Sigma}
\bar{a}_{n-1}(x,s)
\label{4.12}
\end{eqnarray}
Since we are interested  only in the first order  of $(1-\alpha)$ we
may take
$\bar{a}_{n-1}=\bar{a}^{st}_{n-1}$ in the r.h.s. of (\ref{4.12})
neglecting the corresponding $\bar{a}_{n-1,\alpha}$ term.
One can see that   $a_n$ has the same  volume part
$\bar{a}^{st}_n$ as  (\ref{4.9})  (see (\ref{B3})):

\begin{eqnarray}
&&a_0^{st}(x)=1 \ \ , \ \ a^{st}_1=({1 \over 6}-\xi ) \bar{R}
\nonumber \\
&&a^{st}_2 (x)={1 \over 180} \bar{R}^2_{\mu\nu\alpha\beta} -
{1 \over 180} \bar{R}^2_{\mu\nu} -{1 \over 6} ({1 \over 5}-\xi)\Box
\bar{R}
+{1 \over 2}({1 \over 6}-\xi )^2 \bar{R}^2
\label{4.13}
\end{eqnarray}
The difference appears in the surface term. For the few first
coefficients we obtain (cf. (\ref{B4})):

\begin{eqnarray}
&& a_{0, \alpha}=0, \ \ a_{1,\alpha}=4\pi (1-\alpha ) \left( {1 \over 6}
({1
+\alpha \over 2\alpha} )
-\xi \right), \nonumber \\
&&a_{2, \alpha} = 4\pi (1-\alpha ) ({1 \over 6}-\xi ) ({1 \over 6} ({1
+\alpha
\over 2\alpha})
-\xi  ) \bar{R} - {\pi \over 180} ({1-\alpha^4 \over  \alpha^3})
(\bar{R}_{ij}-
2\bar{R}_{ijij} ) ~~~,
\label{4.14}
\end{eqnarray}
where $\bar{R}_{ii}=\bar{R}_{\mu\nu}n_i^{\mu}n_i^{\nu}$ and
$\bar{R}_{ijij}=\bar{R}_{\mu\nu\lambda\rho}n^{\mu}_in^{\lambda}_
i
n^{\nu}_j n^{\rho}_j $.

\bigskip

Now we are ready to calculate the divergences of the effective
action
$W_{eff}=- \ln Z$. In four dimensions  the infinite part of the
effective
action is the following

\begin{equation}
W_{div}=-{1 \over 32 \pi^2} ({1 \over 2} {a_0 \over \epsilon^4}+{a_1
\over
\epsilon^2} +2a_2 \ln {L \over \epsilon } ),
\label{4.15}
\end{equation}
where $L$ is infrared cut-off. Due to the same property (\ref{4.9})
of  the
coefficients
$a_n$ (\ref{4.12}) the $W_{div}$ is a sum of volume  and surface
parts
(\ref{3.1}).
Combining volume part of the one-loop action (\ref{4.15}) with the
tree-level
one (\ref{3.2}) we can see that divergences  (under $\epsilon
\rightarrow 0$)
are absorbed in the standard renormalization of the coupling
constants
\cite{BD}:

\begin{eqnarray}
&&{1 \over G_{ren}}={1 \over G_{B}}+ {1 \over 2\pi \epsilon^2}
({1 \over 6} -\xi), \ \ c_{1,ren}=c_{1,B}-{1 \over 32 \pi^2} ({1 \over 6}
-\xi)^2 \ln
{L \over \epsilon} \nonumber \\
&&c_{2,ren}=c_{2,B} +{1 \over 32 \pi^2} {1 \over 90} \ln {L \over
\epsilon}
; \ \ c_{3,ren}=c_{3,B} -{1 \over 32 \pi^2} {1 \over 90} \ln {L \over
\epsilon}
\label{4.16}
\end{eqnarray}

On the other hand, applying  the formula

$$
S_{div}=(\alpha \partial_\alpha -1) W_{div}|_{\alpha=1}~~,
$$

we obtain  the divergence of the entropy

\begin{equation}
S_{div}= {1 \over 8\pi \epsilon^2} ({1 \over 6} -\xi) A_\Sigma + \left(
{1
\over 4 \pi}
({1 \over 6} -\xi)^2  \int_\Sigma \bar{R}
-{1 \over 16 \pi } {1 \over 45} \int_{\Sigma}
(\bar{R}_{ii}-2\bar{R}_{ijij} )
\right)
\ln {L \over \epsilon}
\label{4.17}
\end{equation}
We see that the complete entropy which is sum of the tree-level
$S(G_B,c_{i,B})$ (\ref{3.4}) and $S_{div}(\epsilon ) $ (\ref{4.17})
becomes finite by the same renormalization of  the constants
(\ref{4.16})
which renormalizes the effective action. So the identity (\ref{3.5})
indeed holds.

For the minimal coupling  $(\xi=0)$ the expression (\ref{4.17})  has
been
obtained in
\cite{Sol1}. In the conformal invariant case ($\xi={1 \over 6}$) the
Newton
constant $G$ and  the coupling $c_1$ are not renormalized.
Correspondingly,
there are no area $A_\Sigma$ and $\int_{\Sigma}R$ contributions
to the entropy
(\ref{4.17}) which is remarkably determined  by  only  conformally
invariant
 expression $\int_{\Sigma} (R_{ii}-2R_{ijij})$.

It should be noted that  our proof of the main statement is based
on the nice
property of  the heat kernel coefficients $a_n$.  Namely,  up to
$(1-\alpha)^2$
terms the exact
$a_n$  on manifold  $E_\alpha$
occurs to be equal to the standard volume coefficient
$\bar{a}_n^{st}$
 considered on manifold $E_\alpha$:

\begin{equation}
a_n (E_\alpha )=  \int_{E_\alpha} \bar{a}^{st}_n (x,x) +
O((1-\alpha)^2)
\label{4.18}
\end{equation}
if  one applies the formulas of Appendix A. for curvatures on
$E_\alpha$.
Then up to  $(1-\alpha)^2$ the renormalization of entropy
(\ref{3.5})
directly follows from the renormalization of the effective action
(\ref{3.6}).

The curvature terms enter  the matter action of  the fields of
different spins
that gives rise to difficulties  in operating with entanglement
entropy
\cite{Kabat}.  We believe that our result can be certainly
generalized also for
these cases.

\bigskip

\section{Remarks.}
\setcounter{equation}0
One can look  at the entanglement entropy given by  the
expression  (\ref{2.6})
from quite different point of view. Consider the whole system
(gravity plus matter) at arbitrary temperature $T=(2\pi\beta)^{-1}$.
Then its
partition function is given by the Euclidean functional integral
over all
fields
defined on manifold with abelian isometry along the Killing vector
$\partial_\varphi$. They are periodic  with period $2\pi\beta$.
Assumption that
the system includes black hole means that there exists a surface
$\Sigma$
(horizon) which is a fixed point of the isometry. Semiclassically,
we take a
metric satisfying these conditions and evaluate the quantum
contribution of
matter fields on this background. Then  (\ref{2.5}) and (\ref{2.7})
are exactly
such the partition function with  the effective action $W(\beta,
g_{\mu\nu})=-
\ln Z$ to be
 the  functional of the temperature $\beta^{-1}$ and  the metric
$g_{\mu\nu}$.
Taking its variation with respect to $\beta$ (when $g_{\mu\nu}$
fixed) gives us
the  statistical (entanglement) entropy
$S_{ent}=(\beta \partial_\beta -1)W(\beta , g_{\mu\nu})$ above
considered.

On the other hand, taking  temperature to be fixed we can find the
corresponding equilibrium
configuration which is  extremum of the effective action $W(\beta,
g_{\mu\nu})$.
The entanglement entropy  then is worth comparing  with the
thermodynamical
entropy\footnote{ I wish to thank V.P.Frolov for discussing this
point.} of a
black hole. The latter  is determined  by total response of the
one-loop
free energy $F$ ($\beta F= W$) of the system being in thermal
equilibrium on
variation of temperature.
 So we must compare the free energies of two  configurations
being  in
equilibrium at slightly different
temperatures.
 The equilibrium configuration corresponding to the fixed
temperature $\beta$
is found from the extreme
 equation ${\delta W (\beta , g_{\mu\nu}) \over \delta
g_{\mu\nu}}|_{\beta
}=0$. This extremum of the effective action is reached
on regular manifolds without conical singularities and the
equilibrium metric
is  function of  the temperature $\beta$ and parameters fixing the
macro-state
of the system (mass $M$, charge $Q$, etc.)\footnote{Really the
minimization of
the functional $W(\beta, g_{\mu\nu})$ under $\beta$ fixed
includes also
variations in the space of macro-parameters. Therefore,  the
equilibrium state
lies on the constraint  $\beta=\beta (M, Q)$. }.
Now the equilibrium free energy $\beta F=W(\beta ,
g_{\mu\nu}(\beta))$
gives us the thermodynamical entropy $S^{TD}=(\beta  d_\beta-1)
W(\beta, g_{\mu\nu}(\beta))$.  Note that  for equilibrium states the
total
derivative
$d_\beta W=\partial_\beta W +{\delta W (\beta , g_{\mu\nu}) \over
\delta
g_{\mu\nu}} {\delta g_{\mu\nu} \over \delta \beta }$ coincides with
the
partial.  Then  we obtain  that two entropies indeed  coincide,
$S^{TD}=S_{ent}$.

However,  in order to calculate $S^{TD}$ we must know exactly
the form of the
quantum-corrected configuration $g_{\mu\nu}(\beta)$ that
is  normally out of our  knowledge. On the other hand the
calculation
of $S_{ent}$ does not require such an  information and we can
obtain exactly
the entropy  (off-shell) as  a function of metric and its derivatives
on the
horizon
$\Sigma$. It should be emphasized that there is no contribution to
$W(\beta,
g_{\mu\nu}(\beta))$ due to the conical singularity and we deal
with  the
standard ultraviolet divergences coming from the bulk terms in
the effective
action.
They  result  in the corresponding divergences of the entropy
which are
obviously regularized by  the standard renormalization of the
gravitational
couplings.
So in terms of the thermodynamical entropy our main statement
holds
automatically.


\bigskip\bigskip
\begin{center}
{\bf Acknowledgments}
\end{center}
I  am grateful to R.C.Myers, V.P.Frolov , D.V.Fursaev and
A.Zelnikov for
valuable
discussions.  I thank V.P.Frolov for kind hospitality at University
of Alberta
and financial support.

\newpage
{\appendix \noindent{\large \bf Appendix  A: Curvature tensors on
$E_\alpha$
\cite{FS2}.}\\
\def\theequation{A.\arabic{equation}}
\setcounter{equation}0
Consider  space $E_\alpha$ which is $\alpha$-fold covering of
a smooth manifold $E$ along the Killing vector $\partial_\varphi$
generating
abelian isometry. Let  surface $\Sigma$ be a stationary point of
this isometry
and  near $\Sigma$  space $E_\alpha$ looks as direct
product  $\Sigma \times {\cal C}_\alpha$ of the surface $\Sigma$
and
two-dimensional cone ${\cal C}_\alpha$ with angle deficit $\delta=
2\pi(1-\alpha)$.  Outside the singular surface $\Sigma$ the space
$E_\alpha$ has the same geometry as  smooth manifold $E$.
In particular, their curvature tensors  coincide.  However, at the
surface $\Sigma$ there exists a conical singularity which results
in singular (delta-function like) contribution to the curvatures. To
extract
this contribution exactly one can use some regularization
procedure
replacing the singular space $E_\alpha$ by  a sequence  of
regular
manifolds $\tilde{E_\alpha}$. In the limit  $\tilde{E_\alpha}
\rightarrow
E_\alpha$ we obtain the following result \cite{FS2}:

\begin{eqnarray}
&&R^{\mu\nu}_{\ \ \alpha\beta} = \bar{R}^{\mu\nu}_{\ \
\alpha\beta}+
2\pi (1-\alpha) \left( (n^\mu n_\alpha)(n^\nu n_\beta)- (n^\mu
n_\beta)
(n^\nu n_\alpha) \right) \delta_\Sigma \nonumber \\
&&R^{\mu}_{ \ \nu} = \bar{R}^{\mu}_{ \ \nu}+2\pi(1-\alpha)(n^\mu
n_\nu)
\delta_\Sigma \nonumber                            \\
&&R = \bar{R}+4\pi(1-\alpha) \delta_\Sigma
\label{A0}
\end{eqnarray}
where $\delta_\Sigma$ is the delta-function: $\int_{\cal M}^{}f
\delta_\Sigma=
\int_{\Sigma}^{}f$; $n^k=n^k_\mu dx^\mu$ are two orthonormal
vectors
orthogonal to $\Sigma$, $(n_\mu n_\nu)=\sum_{k=1}^{2}n^k_\mu
n^k_\nu$
and the quantities $\bar{R}^{\mu\nu}_{\ \ \alpha\beta}$,
$\bar{R}^{\mu}_{ \
\nu}$ and $\bar{R}$
are computed in the regular points $E_{\alpha}/\Sigma$ by the
standard method.

These formulas can be applied to define the integral expressions:

\begin{equation}
\int_{E_{\alpha}}R=\alpha\int_{E} \bar{R}
+4\pi(1-\alpha)\int_{\Sigma}~~~,
\label{A1}
\end{equation}
\begin{equation}
\int_{E_{\alpha}}R^2=\alpha\int_{E} \bar{R}^2
+8\pi(1-\alpha)\int_{\Sigma}\bar{R}+O((1-\alpha)^2)~~~,
\label{A2}
\end{equation}
\begin{equation}
\int_{E_{\alpha}}R^{\mu\nu}R_{\mu\nu}=\alpha\int_{ E}
\bar{R}^{\mu\nu}\bar{R}_{\mu\nu}
+4\pi(1-\alpha)\int_{\Sigma}\bar{R}_{ii}+O((1-\alpha)^2)~~~,
\label{A3}
\end{equation}
\begin{equation}
\int_{E_{\alpha}}R^{\mu\nu\lambda\rho}R_{\mu\nu\lambda\rho}
=\alpha\int_{E}
\bar{R}^{\mu\nu\lambda\rho}\bar{R}_{\mu\nu\lambda\rho}
+8\pi(1-\alpha)\int_{\Sigma} \bar{R}_{ijij}
+O((1-\alpha)^2)~~~,
\label{A4}
\end{equation}
where $\bar{R}_{ii}=\bar{R}_{\mu\nu}n_i^{\mu}n_i^{\nu}$ and
$\bar{R}_{ijij}=\bar{R}_{\mu\nu\lambda\rho}n^{\mu}_in^{\lambda}_
i
n^{\nu}_j n^{\rho}_j $.

 The first integrals in right part of
(\ref{A1})-(\ref{A4})
are defined on the smooth space $E$; they are proportional to
$\alpha$.
The terms $O((1-\alpha)^2)$ in (\ref{A2})-({\ref{A4}) are really
something like $\delta^2_\Sigma$ . They  are ill-defined and turn
to be
dependent on
the regularization prescription and singular
in the limit ${\tilde E}_{\beta} \rightarrow{\cal M}_{\beta}$. But
these terms are not important, for example, in calculation of
entropy.

\bigskip

{\appendix \noindent{\large \bf Appendix  B:
The heat kernel expansion of operator $(-\Box+\xi \bar{R})$ on
$E_\alpha$
\cite{a4}.} \\
\def\theequation{B.\arabic{equation}}
\setcounter{equation}0
Consider  on  space $E_\alpha$, possessing an abelian isometry,
the operator
$-\Box_\xi =-\Box+ \xi \bar{R}$ ,
where $\bar{R}$ is regular part of the scalar curvature $R$ on
$E_\alpha$
(see (\ref{A0})).  Then  we have  the following heat kernel
expansion

\begin{eqnarray}
&& \ln det (-\Box_\xi )= \int_{\epsilon^2}^{\infty} {ds \over s}  Tr
\bar
{K}_{E_\alpha} (s),
\nonumber \\
&&\bar{K}_{E_\alpha} (s)=e^{-s \Box_\xi}={1 \over (4\pi s)^{d \over
2}}
\sum_{n}^{}{} \bar{a}_n s^n, \ \ s \rightarrow 0
\label{B1}
\end{eqnarray}
where the  $\bar{a}_n (x,x)$

\begin{equation}
\bar{a}_n(x,x)=\bar{a}^{st}_n (x,x)+ \bar{a}_{n, \alpha} (x,x)
\delta_\Sigma
\label{B2}
\end{equation}
is sum of  the standard coefficient  $\bar{a}^{st}_n (x,x)$ for
smooth
manifold $E$ \cite{BD}:

\begin{eqnarray}
&&a_0^{st}(x)=1 \ \ , \ \ a^{st}_1=({1 \over 6}-\xi ) \bar{R}
\nonumber \\
&&a^{st}_2 (x)={1 \over 180} \bar{R}^2_{\mu\nu\alpha\beta} -
{1 \over 180} \bar{R}^2_{\mu\nu} -{1 \over 6} ({1 \over 5}-\xi)\Box
\bar{R}
+{1 \over 2}({1 \over 6}-\xi )^2 \bar{R}^2
\label{B3}
\end{eqnarray}
and  a part coming from the singular surface $\Sigma$ (stationary
point of the
isometry):

\begin{eqnarray}
&&a_{0, \alpha}=0; \ \ \ a_{1, \alpha }={\pi \over
3}{(1-\alpha)(1+\alpha)
\over
\alpha}
\int_{\Sigma}^{}\sqrt{\gamma} d^2 \theta \ ; \nonumber \\
&&a_{2, \alpha }={\pi \over 3} {(1-\alpha)(1+\alpha) \over \alpha}
\int_{\Sigma}^{}({1 \over 6}- \xi) \bar{R} \sqrt{\gamma} d^2 \theta
\nonumber
\\
&&-{\pi \over 180}
{(1-\alpha)(1+\alpha)(1+\alpha^2) \over \alpha^3}
\int_{\Sigma}^{}(\bar{R}_{\mu\nu}n^\mu_i n^\nu_i
-2\bar{R}_{\mu\nu\alpha\beta}n^\mu_i
n^\alpha_i
n^\nu_j n^\beta_j )\sqrt{\gamma} d^2 \theta
\label{B4}
\end{eqnarray}
where $n^i$ are two vectors orthogonal to surface $\Sigma$
($n^\mu_i n_{j}^\nu
g_{\mu\nu}=
\delta_{ij}$) and $\gamma$ is metric on the surface $\Sigma$.


\newpage

\begin{thebibliography} \\
\bibitem{Bekenstein1}  J.D.Bekenstein, Lett.Nuov.Cim. {\bf 4}, 737
(1972);
Phys.Rev. {\bf D7},  2333  (1973) ; Phys.Rev. {\bf D9},  3292 (1974).
\bibitem{Hawking}  S.W.Hawking, Comm.Math.Phys. {\bf 43}, 199
(1975) .
\bibitem{Bekenstein} J.D. Bekenstein, {\it "Do we understand
black
hole entropy?"}, qr-qc/9409015.
\bibitem{a1} G.'t Hooft, Nucl. Phys. {\bf B256}, 727 (1985).
\bibitem{Srednicki} M. Srednicki, Phys. Rev. Lett. {\bf 71}, 666
(1993).
\bibitem{Sorkin} L. Bombelli, R. Koul, J. Lee, and R. Sorkin,
Phys. Rev. {\bf D34}, 373 (1986)
\bibitem{Frol.}  V.Frolov, I.Novikov, Phys.Rev. {\bf D48},  4545
(1993).
\bibitem{CW} C. Callan, F. Wilczek,  Phys.Lett. {\bf B 333}, 55
(1994).
\bibitem{KS} D.Kabat, M.Strassler, Phys.Lett. {\bf B329}, 46 (1994).
\bibitem{Sus} L. Susskind and J. Uglum, Phys. Rev. {\bf D50},
2700 (1994).
\bibitem{Jac} T. Jacobson, {\it Black hole entropy and induced
gravity},
preprint 1994, gr-qc/9404039.
\bibitem{Sol}  S.N. Solodukhin,  Phys.Rev. {\bf D51},  609   (1995).
\bibitem{Sol1} S.N. Solodukhin,  Phys. Rev.  {\bf D51}, 618 (1995).
\bibitem{F} D.V. Fursaev, {\it Black-hole thermodynamics and
renormalization}, preprint DSF-32/94, hep-th/9408066.
\bibitem{deAlwis} S.P. de Alwis, N.Ohta, {\it On the entropy of
quantum fields
in black hole backgrounds}, COLO-HEP-347; hep-th/9412027.
\bibitem{Cognola} G.Cognola, L.Vanzo and S.Zerbini, {\it One loop
quantum corrections to the entropy for a four-dimensional eternal
black hole},
UTF-342; hep-th/9502006.
\bibitem{FS}D.V.Fursaev, S.N.Solodukhin, {\it On one-loop
renormalization of black hole entropy},   JINR E2-94-462,
hep-th/9412020.
\bibitem{BFZ} A.I. Barvinsky, V.P. Frolov and A.I. Zelnikov,
 Phys. Rev. {\bf D51} , 1741 (1995).
\bibitem{Jac1} T. Jacobson, {\it A note on Hartle-Hawking vacua},
gr-qc/9407022.
\bibitem{a4} D.V.Fursaev, Phys. Lett. {\bf B334}, 53 (1994);
J.S.Dowker, {\it Heat kernels on curved cones}, hep-th/9606002,
Class. Quant. Grav. (1994) to be published.
 \bibitem{Teitelboim}  S. Carlip and C. Teitelboim, {\it The off-shell
black
hole},
gr-qc/9312002;
             C. Teitelboim, {\it Topological roots of black hole entropy},
             preprint, April 1994; M. Ba{\~n}ados, C. Teitelboim and J.
Zanelli, Phys. Rev. Lett.
             {\bf 72}, 957 (1994).
\bibitem{FS2} D.V.Fursaev, S.N.Solodukhin, {\it  On the
Description of the
Riemannian Geometry in the Presence of Conical
  Defects},  JINR E2-95-28; hep-th/9501127.
\bibitem{Wald} R.M. Wald, Phys. Rev. {\bf D48}, R3427 (1993);
               V. Iyer and R.M. Wald, Phys. Rev. {\bf D50}, 846 (1994);
               T.A. Jacobson, G. Kang, and R.C. Myers, Phys. Rev. {\bf
D49},
               6587 (1994); V.Iyer, R.M.Wald, {\it A comparison of
Noether
charge and Euclidean methods for computing the entropy of
stationary black
holes},  gr-qc/9503052.
\bibitem{Myers} J.-G. Demers, R.Lafrance and R.C.Myers, {\it
Black hole entropy
without brick walls}, McGill/95-06; gr-qc/9503003.
\bibitem{Allen}  B.Allen, A.C.Ottewill, Phys.Rev. {\bf D42}, 2669
(1990).
\bibitem{BD} N.D.Birrell, P.C.W.Davies, {\it Quantum Fields in
Curved Space},
(Cambridge University Press, Cambridge, 1982).
\bibitem{Kabat}  D.Kabat, {\it Black hole entropy and entropy of
entanglement},
RU-95-06, hep-th/9503016.

\end{thebibliography}
\end{document}